\documentclass[aps,prb,twocolumn,epsfig,amsmath,showpacs]{revtex4-1}

\usepackage{epsfig}
\usepackage{dcolumn}
\usepackage{bm}
\usepackage{color}
\usepackage{multirow}

\begin{document}
\title{Comparative study of high-T$_c$ superconductivity in H$_3$S and H$_3$P}

\author{Hyungju Oh}
\email[Email:\ ]{xtom97@gmail.com}
\author{Sinisa Coh}
\author{Marvin L. Cohen}
\affiliation{Department of Physics, University of California at Berkeley 
and Materials Sciences Division, Lawrence Berkeley National Laboratory, 
Berkeley, California 94720, USA}

\date{\today}

\begin{abstract}
We report on a comparative study of the electronic structure, phonon spectra, and superconducting 
properties for recently discovered superconducting hydrides, H$_3$S and H$_3$P. 
While the electronic structures of these two materials are similar, there are notable changes 
in the phonon spectra and electron-phonon coupling. The low-frequency bond-bending 
modes are softened in H$_3$P and their coupling to the electrons at the Fermi surface 
is enhanced relative to H$_3$S. Nevertheless, coupling to the high-frequency modes is 
reduced so the resulting calculated superconducting transition temperature is reduced from 
$\sim$166~K in H$_3$S to $\sim$76~K in H$_3$P.
\end{abstract}


% 71.15.Mb Density functional theory, local density approximation, gradient and other corrections 
% 71.20.-b Electron density of states and band structure of crystalline solids 
% 74.20.Fg BCS theory and its development
% 74.62.Bf Effects of material synthesis, crystal structure, and chemical composition  

\maketitle

\section{Introduction}

Many materials have been proposed theoretically as conventional phonon-mediated 
superconductors having a high superconducting transition temperature (T$_c$).
Based on the BCS theory\cite{bardeen}, materials with light masses and strong bonds
are promising candidates for high-T$_c$ superconductors\cite{moussa,cohen}
because T$_c$ is scaled by the inverse square root of the atomic mass.
Therefore, theoretical studies have been intensively performed focusing on the
compounds consisting of the lightest hydrogen atom. In experiments, on the other hand,
achieving a high-T$_c$ in hydrogen compounds has not been reported yet.
Recently, it is experimentally reported that, under extreme high pressures of 
100--200 GPa, sulfur hydride transforms to a metallic state and
shows extremely high-T$_c$ up to $\sim$200K\cite{drozdov,troyan}.

To find out the crystal structure of the high-T$_c$ sulfur hydride,
many {\it ab-initio} studies have been done and most of these studies 
have concluded that cubic H$_3$S will form with a H-rich decomposition environment 
under high pressure\cite{duan,wang,errea,bernstein,duan2,li,errea2}.
Furthermore, from electron-phonon coupling (EPC) calculations
\cite{duan,wang,errea,bernstein,duan2,li,errea2,papa,heil,sano},
it is revealed that strong coupling happens between high-frequency phonon modes
and electrons and these strong coupling induces high-T$_c$ 
in the body-centered cubic H$_3$S. 

\begin{figure} % FIG. 1
\epsfig{file=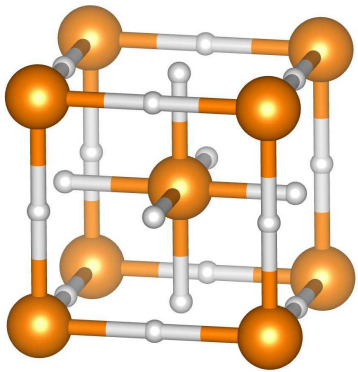,width=5cm,clip=}
\caption{The {\it Im$\overline{3}$m} crystal structure assumed for H$_3$S and H$_3$P.
The large sphere (orange) is S or P, and the small sphere (white) is H.
\label{fig:atom}
}
\end{figure}

Here we study two types of hydrides, H$_3$S and H$_3$P.
Following the discovery of high-T$_c$ conventional superconductivity in sulfur hydride,
a hydride phosphine (H$_3$P) was also reported to be a possible high-T$_c$ (T$_c>100~$K 
at pressure P$>200~$GPa) superconductor via four-probe electrical measurements \cite{drozdov2}.
Hence we compare the normal and superconducting properties of these two materials. 
For the crystal structures of high-T$_c$ hydrides, X-ray diffraction experiments\cite{errea2,einaga} 
confirm that the sulfur atoms of H$_3$S form a body-centered cubic structure 
as shown in Fig.~\ref{fig:atom}. Up to now, no available experimental data for 
the crystal structure of H$_3$P exists. Hence for comparison purposes, 
we assume in this study that both materials have the same crystal structure 
and analyze the effect of element change on material properties.

\section{Methods}

The following methods are used to perform the calculations of the electronic structures, 
phonon properties, and superconducting properties. For the electronic structures,
our calculations are based on {\it ab-initio} norm-conserving
pseudopotentials and the Perdew-Burke-Ernzerhof\cite{perdew}
functional as implemented in the SIESTA\cite{sanchez} and 
Quantum-ESPRESSO\cite{giannozzi} codes.
Phonon frequencies are computed using density-functional perturbation theory\cite{baroni}
implemented in Quantum-ESPRESSO\cite{giannozzi} package.
Finally, EPC and Eliashberg spectral functions are
obtained via the Wannier90\cite{mostofi} and EPW\cite{noffsinger} packages.

For the calculation using SIESTA, electronic wavefunctions are expanded 
with pseudoatomic orbitals (double-$\zeta$ polarization) and a charge density 
cutoff of 800 Ry is used. We sample the Brillouin zone on a uniform 
16$\times$16$\times$16 k-point mesh. For the calculation with Quantum-ESPRESSO,
a plane-wave basis up to 160 Ry and a 32$\times$32$\times$32 k mesh size are employed.

Phonon frequencies $\omega_{{\bf q}\nu}$ and EPC parameters $\lambda_{{\bf q}\nu}$
are computed on a coarse mesh (8$\times$8$\times$8) of reciprocal space.
Next, interpolation techniques\cite{giustino} based on maximally localized Wannier 
functions\cite{giustino,marzari,souza} are used to interpolate EPC 
parameters on a fine grid (36$\times$36$\times$36).

The Eliashberg spectral function $\alpha^2 F(\omega)$ is computed by 
integrating the interpolated phonon frequencies $\omega_{{\bf q}\nu}$ 
and the EPC $\lambda_{{\bf q}\nu}$ over the Brillouin zone,
\begin{equation}
 \alpha^2 F(\omega)=\frac{1}{2} \sum_{{\bf q}\nu} {\mathrm w}_{\bf q} \omega_{{\bf q}\nu} 
 \lambda_{{\bf q}\nu} \delta(\omega-\omega_{{\bf q}\nu}).
 \label{eq:a2f}
\end{equation}
Here the ${\mathrm w}_{\bf q}$ is the Brillouin zone weight 
associated with the phonon wavevectors ${\bf q}$.
The total EPC $\lambda$ is calculated as the Brillouin zone 
average of the mode-resolved coupling strengths $\lambda_{{\bf q}\nu}$:
\begin{equation}
 \lambda=\sum_{{\bf q}\nu} {\mathrm w}_{\bf q} \lambda_{{\bf q}\nu}
 =2 \int_{0}^{\infty} d\omega~\alpha^2 F(\omega) / \omega.
\end{equation}

\section{Electronic structure}

Here we discuss the electronic structure of H$_3$S and H$_3$P. 
In all of our calculations we set the conventional lattice parameter as
3 \AA. With this lattice parameter, the calculated pressures of 
both materials are 220 GPa. 

\begin{figure} % FIG. 2
\epsfig{file=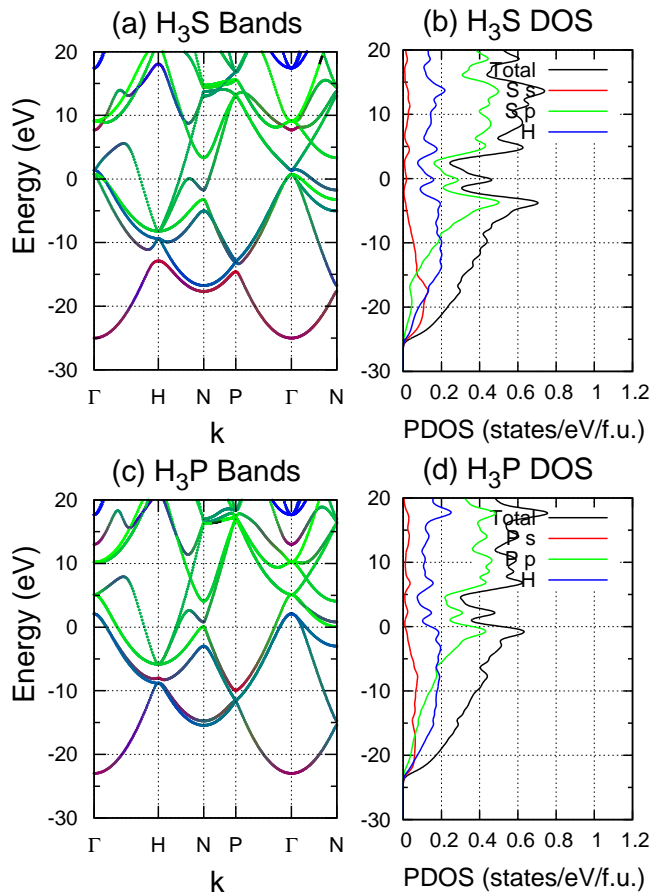,width=8.5cm,clip=}
\caption{Electronic band structures and density of states (DOS) 
per three-hydrogen formula unit (f.u.) of (a), (b) H$_3$S and (c), (d) H$_3$P. 
Dominant orbital characters are represented in blue (H orbitals), 
red (S or P $s$ orbitals), and green (S or P $p$ orbitals) color. 
\label{fig:band}
}
\end{figure}

The overall shapes of the band structures are similar for both materials
[Figs.~\ref{fig:band}(a) and (c)]. Because phosphorus has one less valence 
electron than sulfur, the Fermi level ($E_F$) is shifted down in H$_3$P. 
With the shift, $E_F$ of H$_3$P is placed near a different peak position in the
density of states (DOS). For H$_3$S, the DOS at $E_F$ is calculated to be
0.45~states~eV$^{-1}$~f.u.$^{-1}$. A similar value
(0.50~states~eV$^{-1}$~f.u.$^{-1}$) for the DOS is found in the case of H$_3$P.

Figure~\ref{fig:band} compares the orbital contributions to the band structure and
DOS in H$_3$S and H$_3$P. In both materials, the DOS at $E_F$ comes
dominantly from $3p$ orbitals of sulfur or phosphorus. The portion of $3p$ orbitals 
is twice as large as the portion of hydrogen orbitals.
The Fermi surfaces originated from hydrogen orbitals are almost same in
both case, forming small hole pockets centered at ${\it \Gamma}$-point.

\section{Phonon properties}

\begin{figure} % FIG. 3
\epsfig{file=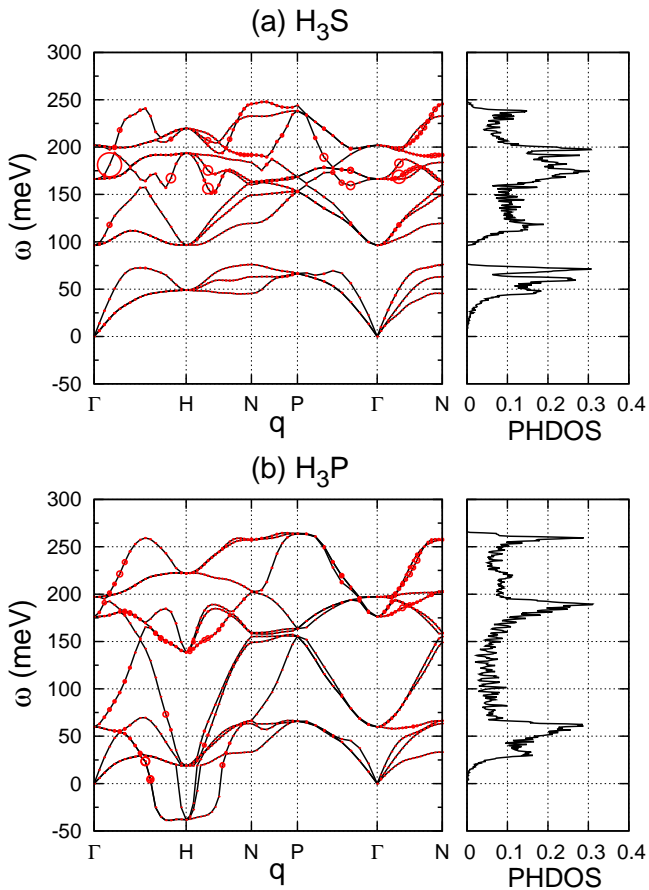,width=8.5cm,angle=0,clip=} 
\caption{
Phonon spectrum and phonon density of states (PHDOS) of 
(a) H$_3$S and (b) H$_3$P. The radius of the red circle is proportional 
to $\omega_{{\bf q}\nu}\lambda_{{\bf q}\nu}$.
\label{fig:phonon}
}
\end{figure}

In this section we discuss the differences of the phonon properties 
between H$_3$S and H$_3$P. When the sulfur is changed to phosphorus, 
the characteristics of the phonon spectra differ significantly along 
${\it \Gamma}$--{\it H} and {\it H}--{\it N} high-symmetry lines [Fig.~\ref{fig:phonon}]. 
The hydrogen--phosphorus bond-bending modes become softer 
and three unstable phonon modes appear at the {\it H} high-symmetry point. 
Therefore we expect that in the doubled unit cell these unstable modes 
would be stabilized. We exclude these negative phonon modes 
when calculating $\alpha^2 F$ so that we can make an reliable 
comparison with H$_3$S. The structural instability of body-centered cubic 
H$_3$P is also reported by previous theoretical structural studies\cite{shamp,liu}.

Next we discuss the strength of the EPC for the two cases. In H$_3$S, 
phonon modes of 150$\sim$200 meV frequencies (which are H--S bond-stretching 
modes) are strongly coupled to electrons at the Fermi surface. 
In H$_3$P, however, low-frequency modes ($<$~50~meV) are more relevant. 
These modes originate from softened H--P bond-bending motion.

To give a more quantitative discussion about the relevant energy scales
of the phonons, we calculate the EPC-weighted average of the phonon frequencies,
\begin{equation}
 \omega_{\mathrm{ln}}=\mathrm{exp} \left\{ \frac{2}{\lambda} \int_{0}^{\infty} 
 d\omega~\frac{\alpha^2 F(\omega)}{\omega}~\mathrm{ln}~\omega \right\}.
\end{equation}
The value of $\omega_{\mathrm{ln}}$ is 1580~K (136~meV) for the H$_3$S 
and 610~K (53~meV) for the H$_3$P. Therefore $\omega_{\mathrm{ln}}$ is 
more than twice as large in H$_3$S relative to H$_3$P.

\section{Superconducting properties}

The total EPC $\lambda$ equals 1.38 in H$_3$S, whereas it reaches 
1.66 in H$_3$P. The Eliashberg phonon spectral functions of H$_3$S and H$_3$P 
are quite different. The EPC in H$_3$S is dominated by the phonon modes 
at the zone center ${\it \Gamma}$ point. In H$_3$P, however, 
we observed an overall contribution of different modes to $\lambda$ along 
${\it \Gamma}$--{\it H}--{\it N} directions as shown in Fig.~\ref{fig:phonon}.

\begin{figure} % FIG. 4
\epsfig{file=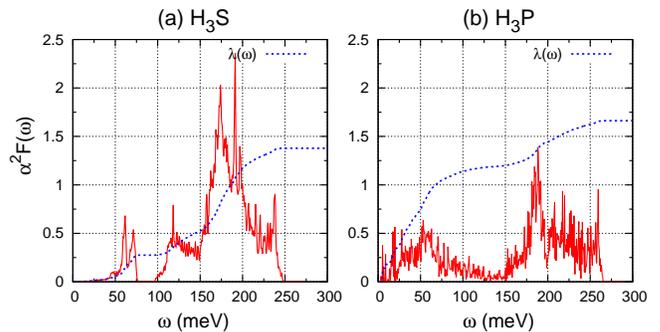,width=8.5cm,angle=0,clip=} 
\caption{
Eliashberg spectral function $\alpha^2F$ (red) and cumulative contribution 
to the electron-phonon coupling strength $\lambda$ (blue) of 
(a) H$_3$S and (b) H$_3$P. The cumulative EPC is calculated as
$ \lambda(\omega)=2 \int_{0}^{\omega} d\omega'~\alpha^2 F(\omega') / \omega'$.
\label{fig:a2f}
}
\end{figure}

Here we discuss why there is a large difference in the EPC 
between H$_3$S and H$_3$P. First, we consider the difference in DOS. 
Since $\lambda$ is roughly proportional to the DOS at $E_F$, the EPC 
could be enhanced by the large DOS. However, in our case, there is 
no sufficient change in DOS to reproduce the large enhancement in EPC for H$_3$P. 
Another point is the coupling strength between the electrons and 
the low-frequency hydrogen vibration. There is no significant enhancement
in the electron-phonon matrix elements which is proportional to 
 $\omega_{{\bf q}\nu}\lambda_{{\bf q}\nu}$ [Fig.~\ref{fig:phonon}].
 But, the dominant modes to EPC appear at low frequencies in H$_3$P [Fig.~\ref{fig:a2f}].
 This change causes the enhancement of the EPC $\lambda$ value.

Finally, we estimate the superconducting transition temperature T$_c$
using the McMillan equation\cite{allen}
\begin{equation}
 T_c = \frac{\omega_{\mathrm{ln}}}{1.20}~\mathrm{exp} \left\{ - \frac
 {1.04~(1+\lambda)}{\lambda-\mu^*(1+0.62~\lambda)} \right\}.
\end{equation}
Here $\mu^*$ is the Coulomb repulsion parameter.  For commonly used
$\mu^*=0.1$ we estimate T$_c=166$~K for H$_3$S and 76~K for H$_3$P.
The exact value of $\mu^*$ here is not that important since even with
$\mu^*=0$ we get very similar T$_c$ (219 and 96~K).  

The value of $\lambda$ we obtained for H$_3$S and H$_3$P is near the limit
of applicability of the McMillan equation.  However, we find that
the Kresin--Barbee--Cohen model\cite{kresin,bourne}, which is
applicable for large $\lambda$, gives similar estimates for T$_c$.

Although H$_3$P has a higher $\lambda$ value than H$_3$S,
the estimated T$_c$ is about half of that in H$_3$S. 
This agrees well with the experimentally obtained T$_c$ of 
$\sim200$~K in H$_3$S and $\sim100$~K in H$_3$P.
We expect that the deviation here from experiment might occur 
because we ignored unstable phonon modes in our calculation, 
so softening might be overestimated for H$_3$P in the low-frequency regime.

\section{Conclusion}

With the assumption of the same body-centered cubic structure 
and lattice parameter, we compare the electronic, phonon, and 
superconducting properties of H$_3$S and H$_3$P. The results of 
electronic structures show no significant difference, except for a slight 
change in the Fermi level due to the different number of valence electrons. 
However, there are notable changes in phonon spectrum and 
electron-phonon coupling properties. First, there exists phonon softening 
in low-frequency bond-bending modes, and the coupling of these modes 
to electrons near the Fermi surface is enhanced. 
As the dominant frequency regime changes from high to low frequency, 
the superconducting transition temperature is reduced 
from $\sim$166~K in H$_3$S to $\sim$76~K in H$_3$P.

\section*{Acknowledgements}

This work was supported by National Science Foundation Grant
No. DMR15-1508412 (electronic structure calculation) and by the
Director, Office of Science, Office of Basic Energy Sciences,
Materials Sciences and Engineering Division, U.S. Department of Energy
under Contract No. DE-AC02-05CH11231, within the Theory Program 
(phonon and superconducting properties calculations). 
Computational resources have been provided by
the DOE at Lawrence Berkeley National Laboratory's NERSC facility.

\end{document}